\documentclass[twocolumn]{aastex6}
\usepackage{amsmath}

\newcommand{\rmn}{\mathrm}
\newcommand{\beq}{\begin{equation}}
\newcommand{\eeq}{\end{equation}}

\begin{document}
\title{A detection of the integrated Sachs--Wolfe imprint of cosmic superstructures using a matched-filter approach}
\shorttitle{Detecting the ISW imprint of superstructures}
\shortauthors{Nadathur \& Crittenden}

\author{Seshadri Nadathur and Robert Crittenden} 
\affil{Institute of Cosmology and Gravitation, University of Portsmouth, Burnaby Road, Portsmouth PO1 3FX, United Kingdom}
\email{seshadri.nadathur@port.ac.uk}

\begin{abstract}
We present a new method for detection of the integrated Sachs--Wolfe (ISW) imprints of cosmic superstructures on the cosmic microwave background, based on a matched filtering approach. The expected signal-to-noise ratio for this method is comparable to that obtained from the full cross-correlation, and unlike other stacked filtering techniques it is not subject to an \emph{a posteriori} bias. We apply this method to \emph{Planck} CMB data using voids and superclusters identified in the CMASS galaxy data from the Sloan Digital Sky Survey Data Release 12, and measure the ISW amplitude to be $A_\rmn{ISW}=1.64\pm0.53$ relative to the $\Lambda$CDM expectation, corresponding to a $3.1\sigma$ detection. In contrast to some previous measurements of the ISW effect of superstructures, our result is in agreement with the $\Lambda$CDM model. 

\end{abstract}


\section{Introduction} \label{sec:intro}
 
Since the discovery of the accelerated expansion of the Universe \citep{Riess:1998,Perlmutter:1999}, the nature of dark energy has become one of the central puzzles in cosmology. The time evolution of gravitational potentials produces secondary temperature anisotropies in the cosmic microwave background (CMB) via the late-time integrated Sachs--Wolfe (ISW) effect \citep{Sachs:1967}. Because this time evolution can be caused by a cosmological constant $\Lambda$, measurement of the ISW signal is a direct probe of the dynamical effects of dark energy \citep{Crittenden:1996}.

The ISW signal is conventionally measured through a cross-correlation between the CMB and large-scale structure (LSS) tracers \citep[e.g.][]{Fosalba:2003,Afshordi:2004,Boughn:2004,Nolta:2004,Giannantonio:2006,Giannantonio:2008,Ho:2008}. This method yields typical detection significances of 2$\sigma$--3$\sigma$ for individual LSS tracers, rising to $\sim4\sigma$ through a combination of multiple tracers \citep{Giannantonio:2012,Planck:2015ISW}.

An alternative method for detection is to stack filtered CMB patches around the locations of localised `superstructures' -- large empty cosmic voids and overdensities known as superclusters. \citet{Granett:2008a} used this method with WMAP CMB data and 100 superstructures from SDSS to report a $\sim4.5\sigma$ detection, subsequently confirmed with \emph{Planck} data \citep{Planck:2013ISW,Planck:2015ISW}.

However, this detection has proven difficult to interpret due to the large amplitude of the observed effect. Theoretical estimates with optimistic assumptions \citep{Nadathur:2012,Flender:2013,Aiola:2015} showed that it exceeds the maximum expectation in a $\Lambda$CDM cosmology by a factor of $\sim5$, a $\gtrsim3\sigma$ discrepancy with theory. Simulation results give an even smaller expected signal \citep{Cai:2014,Hotchkiss:2015a}, exacerbating the problem. 

Efforts to replicate the measurement with independent superstructure catalogs have either given null results \citep{Ilic:2013,Hotchkiss:2015a} or marginally significant detections with amplitude still in excess of expectation \citep{Cai:2014,Granett:2015,Kovacs:2015}. Such studies often rely on arbitrary choices for the number of superstructures included in the stacks and the width of the compensated top-hat (CTH) filter used in the analysis, potentially introducing important \emph{a posteriori} biases in the analysis \citep{HernandezMonteagudo:2013}.

In this study, we describe a new method for the detection of the stacked ISW signal using matched filters constructed after calibration on simulations. Our method has far higher expected sensitivity, even comparable with that expected from the full cross-correlation technique, and is free of possible \emph{a posteriori} bias. We applied this method to data from the CMASS galaxy sample from the SDSS Data Release 12 (DR12) and \emph{Planck}, and report a detection of the ISW effect of superstructures at $3.1\sigma$ significance. The amplitude of the ISW effect is consistent with $\Lambda$CDM expectations, thus potentially resolving this long-standing anomaly. 
 
\section{Data sets} \label{sec:data}

\subsection{LSS data and simulations} \label{subsec:lss}

We identified cosmic voids and superclusters in the CMASS galaxy sample of the SDSS-III BOSS DR12 \citep{Alam-DR11&12:2015}.\footnote{\url{http://data.sdss3.org/sas/dr12/boss/lss/}} This is the final data release of SDSS-III. The BOSS large-scale structure (LSS) galaxy catalogs provide spectra and redshifts for 1.3 million galaxies over 9,376 deg$^2$ in two contiguous sky regions in the Northern and Southern Galactic Caps. The CMASS sample includes $777\,202$ luminous galaxies in the redshift range $0.43\leq z\leq0.7$ and is selected to be approximately volume-limited in stellar mass. Details on the target selection, data reduction algorithms, and catalog creation are given in \citet{Reid-DR12:2016}. 

To find voids and superclusters we used a modified version of the {\small ZOBOV} algorithm \citep{Neyrinck:2008}, following \citet{Nadathur:2016a}. {\small ZOBOV} reconstructs the local galaxy density field from the discrete galaxy distribution using a Voronoi tessellation, identifies local extrema of the density field, and then uses a watershed algorithm to demarcate individual structures. To prevent the tessellation from leaking beyond the observed survey volume we add a thin layer of buffer particles around the boundary of the survey footprint, within holes in the survey mask, and along both the high- and low-redshift caps \citep[see][]{Nadathur:2016a}. The volumes of the Voronoi cells associated with each galaxy are inverted to estimate local tracer densities, after applying a redshift- and position-dependent weighting to account for variations in the survey mean density $n(z)$ and the survey sky completeness.

Our implementation of the watershed algorithm for void-finding followed that of \citet{Nadathur:2016a}. In particular, we did not merge neighbouring voids together, separating individual structures purely on the basis of the underlying topology of the density field. For superclusters, we applied the same void-finding algorithm to the inverse of the density field, thus identifying density maxima instead of minima. For each superstructure, we determined the average galaxy density contrast, $\overline\delta_g = \frac{1}{V}\int_{V}\delta_g\,\rmn{d}^3\mathbf{x}$, and effective spherical radius, $R_\rmn{eff}= \left(\frac{3}{4\pi}V\right)^{1/3}$, where the superstructure volume $V$ is determined from the sum of the volumes of Voronoi cells making up the structure. The center of each void is defined as the center of the largest completely empty sphere that can be inscribed within it \citep{Nadathur:2015b,Nadathur:2016a}. For superclusters, we took the location of the galaxy with the smallest Voronoi cell within the supercluster as its center.

For calibration of the expected ISW signal from voids and superclusters, we compared the gravitational potential information in the Big MultiDark (BigMD) $N$-body simulation \citep{Klypin:2016} with superstructures found in a mock CMASS galaxy catalog created in that simulation \citep[for details, see][]{Nadathur:2016c}. Previous work using this simulation has shown that such voids and superclusters correspond to large but relatively shallow matter density perturbations within the linear or quasi-linear regime, extending over scales of up to $\mathcal{O}(100\;h^{-1}$Mpc) \citep{Nadathur:2015b,Nadathur:2015c,Nadathur:2016c}. 

To test the operation of our algorithm on CMASS data and to estimate error covariances, we apply the same structure-finding procedure to 1000 sets of mock galaxy catalogs created using the ``quick particle mesh" (QPM) technique \citep{White:2014}. These mocks are based on a set of low-resolution particle mesh simulations consisting of $1280^3$ particles in a box of side $2.56\;h^{-1}$Gpc, with cosmological parameters $\Omega_\rmn{m}=0.29$, $h=0.7$, $n_s=0.97$, and $\sigma_8=0.8$. Halos in the simulations are populated with galaxies using a halo occupation distribution method to reproduce the observed galaxy clustering amplitude. These mocks also incorporate observational effects of the survey selection, veto mask and fibre collisions. The distribution of superstructures in the QPM mocks revealed some localised residual effects of the survey boundary on the tessellation near some of the holes in the survey mask. All superstructures in these regions were treated as contaminated and removed from both the CMASS catalogs and the mocks.

The final catalog of structures used in the analysis has been made available for download.\footnote{\url{http://www.icg.port.ac.uk/stable/nadathur/voids/}}

\begin{figure*}
\centering
\plotone{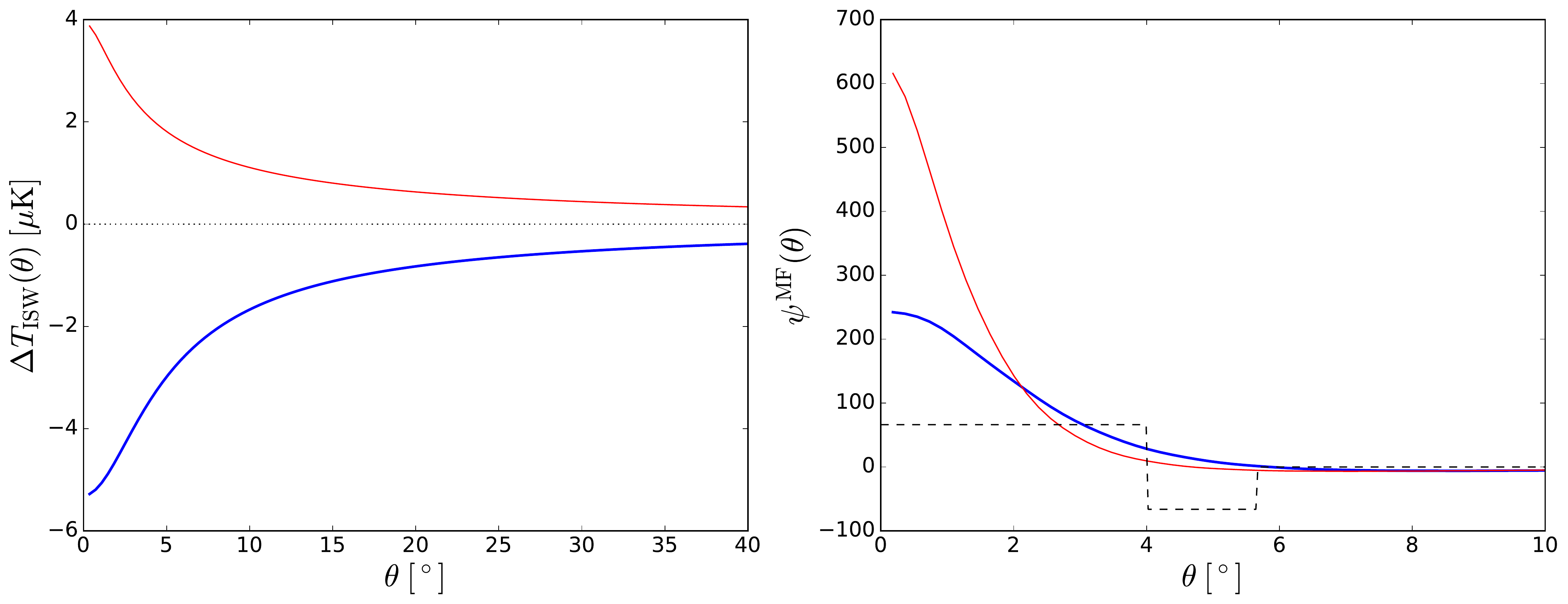}
\caption{Left: template ISW temperature profiles calculated for a void with $\lambda_v=-36.9$ (thick blue line) and a supercluster with $\lambda_c=272.4$ (thin red), both assumed to be centered at redshift $z=0.55$. Right: optimal matched filters constructed for these two templates using equation \ref{eq:MFsphH}. For comparison the dashed black line shows a CTH filter of width $4^\circ$ as used by \citet{Granett:2008a}.
\label{fig:filters}}
\end{figure*}

\subsection{CMB data}

We used the four foreground-cleaned CMB temperature maps from the \emph{Planck} 2015 data release. These are the {\tt\string COMMANDER}, {\tt\string NILC}, {\tt\string SEVEM} and {\tt\string SMICA} maps, named after the component separation methods used to generate them \citep{Planck:2015IX}. The maps are pixelized in {\tt\string HEALPix} format \citep{Gorski:2004by}, at resolution $N_\rmn{side}=1024$, corresponding to a mean pixel spacing of $3.4$ arcminutes. To these maps we applied the common \emph{Planck} UT78 temperature mask, downgraded to the same $N_\rmn{side}$ resolution using a binary threshold cut of 0.9, to eliminate contamination from the Galactic plane and known point sources.

\section{Method} \label{sec:methods}

\subsection{Constructing template profiles} \label{subsec:templates}

The ISW temperature shift along direction $\mathbf{n}$ is given by the line-of-sight integral
\beq
\label{eq:ISWdefn}
\frac{\Delta T_\rmn{ISW}}{\overline{T}}(\boldsymbol{\hat n}) = 2\int_0^{z_\rmn{LS}} \frac{a}{H(z)}\dot\Phi\left(\boldsymbol{\hat n},\chi(z)\right)\,\rmn{d}z\;,
\eeq
where the integral extends to the redshift of last scattering,, $z_\rmn{LS}$. In the linear approximation, density perturbations grow as $\dot\delta = \dot D\delta$, where $D(z)$ is the linear growth function. This can be combined with the Poisson equation for $\Phi$ to obtain
\beq
\label{eq:ISWint}
\frac{\Delta T_\rmn{ISW}}{\overline{T}}(\boldsymbol{\hat n}) = -2\int_0^{z_\rmn{LS}} a(z)\left(1-f(z)\right)\Phi\left(\boldsymbol{\hat n},z\right)\,\rmn{d}z\;,
\eeq
where $f= \frac{\rmn{d}\ln D}{\rmn{d}\ln a}$ is the growth rate. This linear approach is an extremely good approximation on the scales of interest \citep[e.g.,][]{Cai:2010hx,Nadathur:2014b}.

The ISW temperature profile produced by a given structure can be calculated given knowledge of the gravitational potential $\Phi(r)$ about its location, which must be determined from calibration with simulation. We follow the results of \citet{Nadathur:2016c}, who   
studied structures identified using a mock galaxy catalog in the BigMD simulation, finding that the value of $\Phi$ at void locations is tightly correlated with the observable quantity 
\beq
\label{eq:lambda_v}
\lambda_v\equiv\overline\delta_g\left(\frac{R_{\rmn{eff},v}}{1\;h^{-1}\rmn{Mpc}}\right)^{1.2}.
\eeq
The majority of voids identified by {\small ZOBOV} correspond to local underdensities within globally overdense regions and thus do not give $\Delta T_\rmn{ISW}<0$. However, those voids with $\lambda_v<0$ are on average undercompensated, corresponding to regions with $\Phi>0$ and thus a negative ISW shift. For such voids, \citet{Nadathur:2016c} find that the spherically averaged potential profile at distance $r$ from the void center follows the two-parameter form 
\beq
\label{eq:void_Phi}
\overline\Phi(r,\lambda_v) = \frac{\Phi_{0v}(\lambda_v)}{1+\left(r/r_{0v}(\lambda_v)\right)^2}\;,
\eeq
with $\Phi_{0v}(\lambda_v)$ and $r_{0v}(\lambda_v)$ calibrated from fits to simulation.

We examined the properties of superclusters in the same BigMD simulation and found an analogous result. For superclusters, the observable 
\beq
\label{eq:lambda_c}
\lambda_c\equiv\overline\delta_g\left(\frac{R_{\rmn{eff},c}}{1\;h^{-1}\rmn{Mpc}}\right)^{1.6}
\eeq
is an excellent empirical predictor for the value of $\Phi$. Superclusters with $\lambda_c>0$ on average correspond to $\Phi<0$ and thus $\Delta T_\rmn{ISW}>0$. We found the average potential profile for these structures followed 
\beq
\label{eq:cluster_Phi}
\overline\Phi(r,\lambda_c) = \frac{\Phi_{0c}(\lambda_c)}{1+\left(r/r_{0c}(\lambda_c)\right)^{\alpha(\lambda_c)}}\;,
\eeq
with $\Phi_{0c}(\lambda_c)$, $r_{0c}(\lambda_c)$ and $\alpha(\lambda_c)$ again determined from simulation. Importantly, the amplitude of the potential fluctuation scales linearly with both observables, $\Phi_{0v,c}\propto-\lambda_{v,c}$ \citep[see also][]{Nadathur:2016c}. The length scales of the potential perturbations, $r_{0v}$ and $r_{0c}$, typically far exceed the physical extents of the voids and superclusters, as expected from the Poisson equation.

We used the fitted profile forms from equations~\ref{eq:void_Phi} and \ref{eq:cluster_Phi} with equation~\ref{eq:ISWint} to calculate the expected ISW temperature shift $\Delta T_\rmn{ISW}(\theta)$ at angle $\theta$ from the line of sight to a given void or supercluster, located at a given redshift. This can be split into an amplitude and a spatial profile normalized to unity, as $\Delta T_\rmn{ISW}(\theta) = T_0 y(\theta)$. The axisymmetric template profile can be expanded into spherical harmonics, as
\beq
\label{eq:y_sphH}
y(\theta) = \sum_{\ell=0}^\infty y_{\ell 0} Y_\ell^0(\cos \theta).
\eeq

\subsection{Matched-filter construction} \label{subsec:filters}

The total temperature signal at sky location $\boldsymbol{\theta}=\left(\vartheta,\varphi\right)$ can be written as
\beq
s(\boldsymbol{\theta}) = \Delta T_\rmn{ISW}(|\boldsymbol{\theta}-\boldsymbol{\theta_0}|) + n(\boldsymbol{\theta}),
\eeq
where $\Delta T_\rmn{ISW}$ is the template ISW contribution calculated above, $\boldsymbol\theta_0$ is the location of the structure center, and $n(\boldsymbol{\theta})$ includes all other sources of noise in the foreground-cleaned maps, to which the dominant contribution comes from primordial CMB fluctuations at the last scattering surface. To isolate the ISW signal from the noise, we apply an axisymmetric matched-filter $\psi^\rmn{MF}(\theta)$ to the observed map. The resulting filtered map $u(\boldsymbol\beta)$ is a convolution of the filter function and the observed map,
\beq
\label{eq:filter}
u(\boldsymbol{\beta}) = \int \rmn{d} \Omega\;s(\boldsymbol{\theta})\psi^\rmn{MF}(|\boldsymbol{\theta}-\boldsymbol{\beta}|).
\eeq
The coefficients of the spherical harmonic expansion of the filtered map can be written as \citep{Schaefer:2006}
\beq
\label{eq:filtered_lm}
u_{\ell m} = \sqrt{\frac{4\pi}{2\ell+1}}s_{\ell m}\psi^\rmn{MF}_{\ell 0}.
\eeq

Given a template profile from equation~\ref{eq:y_sphH}, the optimal matched-filter $\psi^\rmn{MF}$ satisfies two constraints:
\begin{enumerate}
\item the expectation value of the filtered field at the structure location is an unbiased estimator of the ISW amplitude, $\langle u(\boldsymbol{\theta_0})\rangle  = T_0$, and
\item the variance of the filtered field is minimized.
\end{enumerate}
These constraints are satisfied by choosing \citep{Schaefer:2006,McEwen:2008}
\beq
\label{eq:MFsphH}
\psi^\rmn{MF}_{\ell 0} = \kappa\frac{y_{\ell 0}}{C_\ell}\,,
\eeq
where $\kappa^{-1}=\sum_{\ell=0}^\infty y_{\ell 0}^2/C_\ell$ and $C_\ell$ denotes the power spectrum multipoles of the noise field -- in this case, the power spectrum of the \emph{Planck} CMB maps. 

We determined the appropriate choice of filter coefficients $\psi^\rmn{MF}_{\ell 0}$ from the template $y_{\ell 0}$ and equation \ref{eq:MFsphH}. The filter depends on the fit parameters $r_{0v}(\lambda_v)$ (for voids) or $r_{0c}(\lambda_c)$ and $\alpha(\lambda_c)$ (for superclusters), and on the comoving distance to the structure center. The signal amplitude $T_0$ depends on all of these as well as on the amplitude of the potential fluctuation $\Phi_{0v,c}$.

Figure~\ref{fig:filters} shows example templates $\Delta T_\rmn{ISW}(\theta)$ and matched-filter profiles $\psi^\rmn{MF}(\theta)$ for a void with $\lambda_v=-36.9$ and a supercluster with $\lambda_c=272.4$. Both are assumed to be centered at redshift $z=0.55$. Also shown for comparison is a CTH filter, used by \citet{Granett:2008a} and other previous ISW stacking analyses.

All of our superstructure templates have expected $S/N\lesssim0.1$. Therefore a detection is only possible by stacking large numbers of structures.

\subsection{Detection strategy} \label{subsec:strategy}

To measure the ISW effect of superstructures we first pruned the structure catalogs to remove all voids with $\lambda_v>0$ and superclusters with $\lambda_c<0$, as they are not expected to contribute an ISW shift of the right sign. This left a total of 2445 voids and 29,866 superclusters. We then binned the remaining structures according to their values of $\lambda_v$ and $\lambda_c$ respectively. 

For each bin, we determined the average $\overline\lambda_v$ (or $\overline\lambda_c$) and the mean redshift of structures in the bin. From these and the fits to equations \ref{eq:void_Phi} and \ref{eq:cluster_Phi} from simulations, we obtained the expected ISW amplitude $T_0$, the profile $y(\theta)$, and the appropriate matched-filter $\psi^\rmn{MF}(\theta)$ for the representative template in each bin. Details of the fits to $\Phi(r)$ used are provided for download together with the superstructure catalogs. 

Note that we constructed the matched filters entirely based on calibration with simulation sand before any reference to the CMB data. No free parameters remain in our analysis, so it is free of any \emph{a posteriori} bias. 

We filtered the appropriately masked CMB maps with the matched filters and stacked the results at the locations of superstructures in each bin to obtain the average values $T_0^\rmn{measured}=\overline{u(\boldsymbol{\theta_0})}$ in each case. To estimate errors we generated 2000 mock measurements by applying the same filters to random realizations of the CMB sky, at the locations of superstructures in randomly selected catalogs from the set of QPM mocks. We also tested generating mocks by fixing the superstructure catalog but still varying the CMB maps, or by varying the catalog but fixing the \emph{Planck} CMB map. These gave comparable results, but do not simultaneously capture both sources of variability.

\begin{figure*}
\centering
\plotone{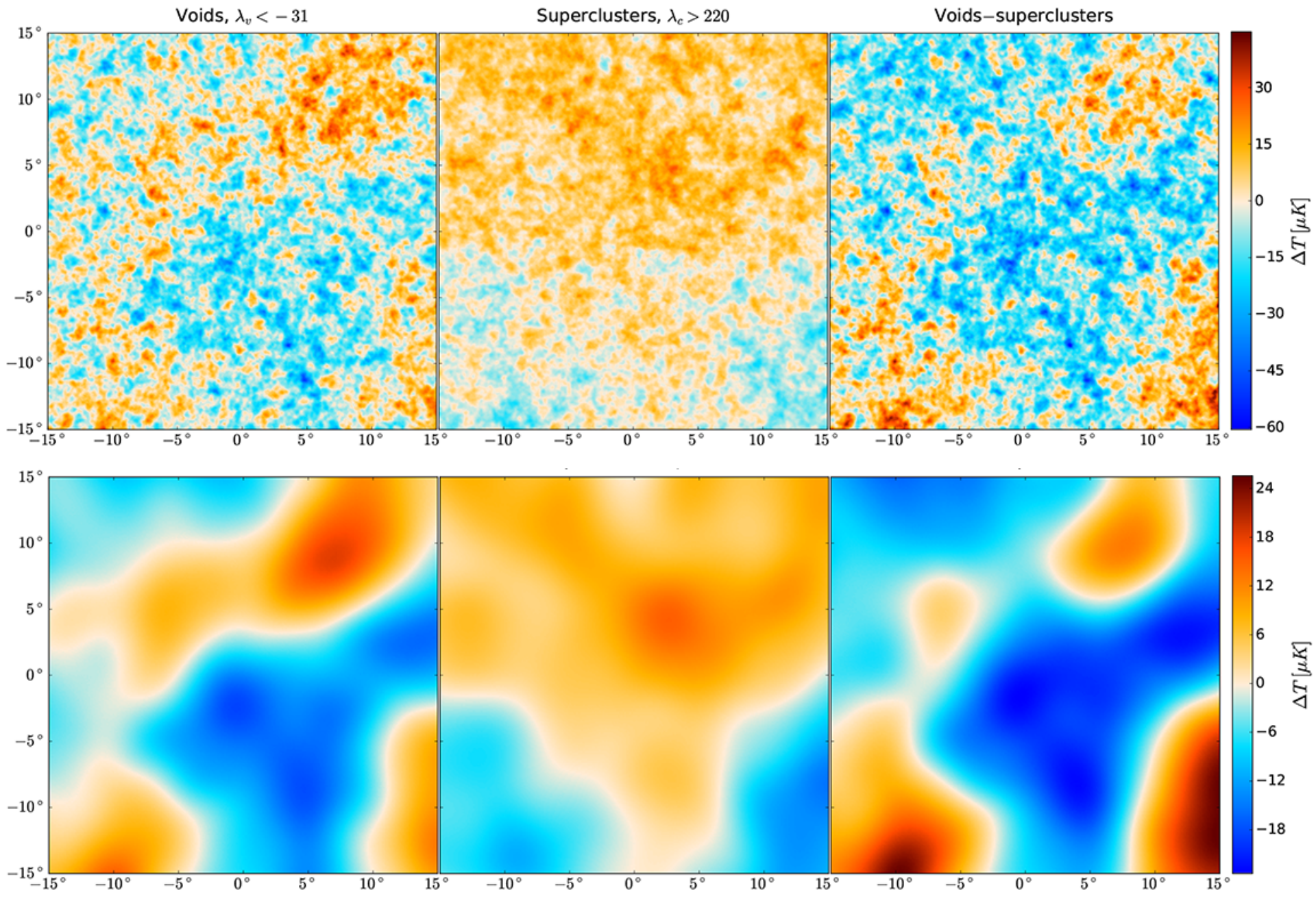}
\caption{\emph{Top row}: Stacked regions of the unfiltered \emph{Planck} {\tt\string SEVEM} map centered on the positions of 96 voids with $\lambda_v<-31$ (left), $353$ superclusters with $\lambda_c>220$ (center), and the difference between them (right). These correspond to the two extreme bins in Figure~\ref{fig:results}, and to the template profiles in Figure~\ref{fig:filters}. \emph{Bottom row}: Same as above, but after filtering with the appropriate matched filters $\psi^\rmn{MF}(\theta)$. 
\label{fig:projections}}
\end{figure*}

\begin{figure*}
\centering
\plotone{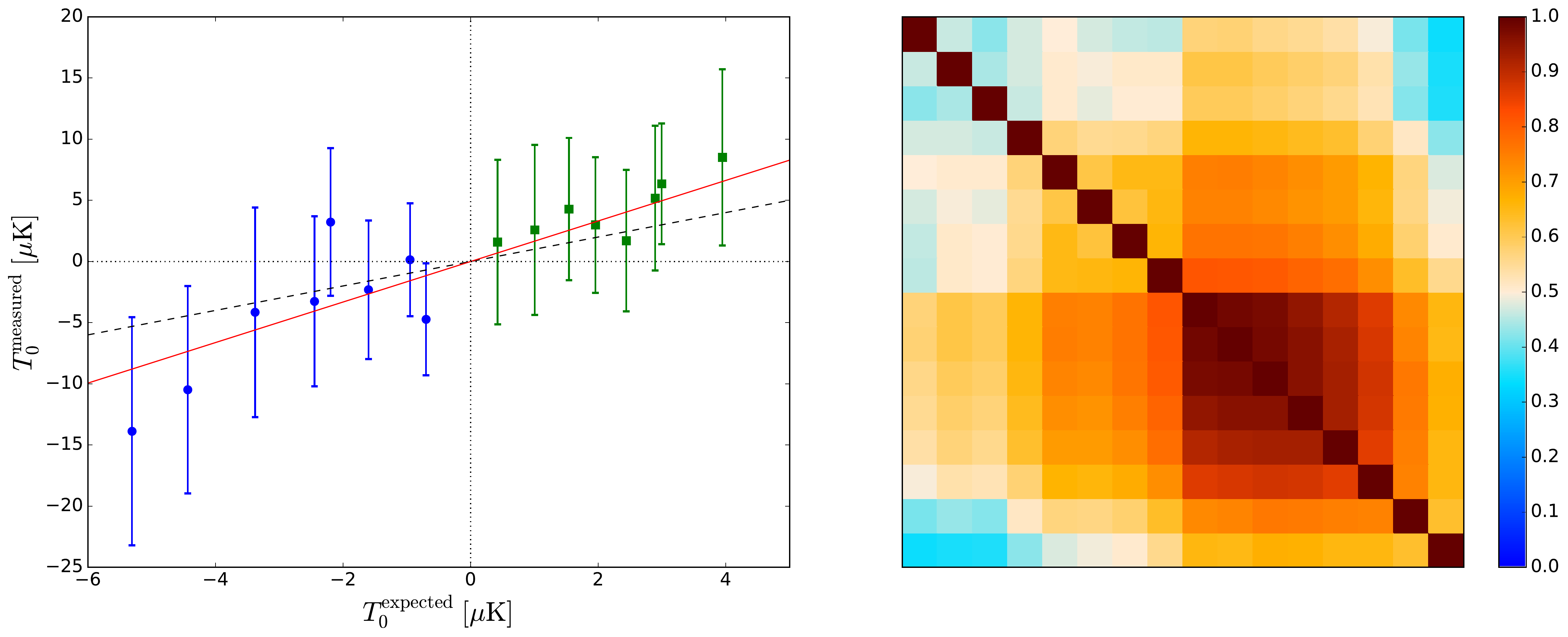}
\caption{Left: mean filtered temperatures at superstructure locations measured on the \emph{Planck} {\tt\string SEVEM} map, as a function of the expected values. Void bins are shown by blue circles and supercluster bins are shown by green squares. Error bars are obtained from diagonal entries of the covariance matrix. The solid red line shows the best-fit value $A_\rmn{ISW}=1.66$ for this map; the black dashed line is $A_\rmn{ISW}=1$. Right: the normalized covariance matrix for these binned measurements determined from mock realizations.
\label{fig:results}}
\end{figure*}

\section{Results} \label{sec:results}

Figure~\ref{fig:projections} shows stacked CMB map patches centered on voids and superclusters in the two extreme bins, containing structures with $\lambda_v<-31$ and $\lambda_c>220$ respectively. A visually compelling difference can be seen between the two stacks, although comparison with the templates in Figure~\ref{fig:filters} shows the difference in magnitude between the ISW signal and the primordial CMB noise, highlighting the need for a statistical approach.

In the left panel of Figure~\ref{fig:results} we show the results obtained for $T_0^\rmn{measured}$ in each of the 16 bins for the {\tt\string SEVEM} map, as a function of the expected ISW amplitude in the bin. Measurements in different bins are highly correlated with each other, as can be seen from the normalized covariance matrix shown in the right panel. We fitted a straight line of the form 
\beq
\label{eq:Aisw}
T_0^\rmn{measured} = A_\rmn{ISW}T_0^\rmn{expected}
\eeq
to this data, where the free parameter $A_\rmn{ISW}=1$ for a $\Lambda$CDM cosmology. The results obtained for each of the four CMB maps are summarized in Table~\ref{table:results}. The likelihood function for $A_\rmn{ISW}$ is Gaussian in all cases.  All four maps give very similar results, indicating a cosmological origin for the signal. A simple average of the results gives $A_\rmn{ISW}=1.64\pm0.53$, with a signal-to-noise ratio of $3.1$.  

For the same structures, using CTH filters with widths in each bin chosen to maximize the available signal from the template profiles gave $A_\rmn{ISW}=1.54\pm0.73$, consistent with our headline result but with $38\%$ larger uncertainty, due to the suboptimal filter choice.

\begin{deluxetable}{ccc}
\tablecaption{Measurements of the ISW Amplitude from Matched-filter Stacking Analysis with \emph{Planck} and CMASS Superstructures. \label{table:results}}
\tablehead{
\colhead{CMB map} & \colhead{$A_\mathrm{ISW}\pm\sigma_A$} & \colhead{$S/N$}}
\startdata
{\tt\string COMMANDER} & $1.65\pm0.53$ & 3.13  \\
{\tt\string NILC} & $1.62\pm0.53$ & 3.07  \\
{\tt\string SEVEM} & $1.66\pm0.53$ & 3.15  \\
{\tt\string SMICA} & $1.62\pm0.53$ & 3.08  \\
\enddata
\end{deluxetable}

\section{Discussion} \label{sec:discussion}

We have presented a new method for detecting the ISW temperature shift in the CMB due to cosmic superstructures, using a combination of stacking and a matched-filter analysis calibrated on simulations. Applying this method to superstructures in the CMASS galaxy data, we obtain a measurement of the ISW amplitude $A_\rmn{ISW}=1.64\pm0.53$, significant at the $3.1\sigma$ equivalent level. This value is insensitive to the method of foreground removal in the CMB maps, pointing to its cosmological origin. This detection significance is among the highest obtained for the ISW using any single LSS tracer \citep{Giannantonio:2012,Planck:2015ISW}.

An important advantage of our new method is that all analysis choices and parameters have been fixed purely based on calibration with the simulation before looking at the CMB data. This means that our measurement is not subject to any \emph{a posteriori} bias. In contrast, several previous ISW stacking measurements on superstructures have included arbitrary choices of the number of superstructures and width of the CTH filter used, potentially affecting the claimed detection significances \citep{HernandezMonteagudo:2013}.

Our method also provides a greatly increased sensitivity over previous stacking analyses, such that we find an expected $S/N$ of 1.9 even for a standard $\Lambda$CDM cosmology. This is due to a combination of factors: better statistics due to the large size of our new catalog of superstructures; an improved calibration with simulation allowing for a more optimal binning in $\lambda_v$ and $\lambda_c$; and the use of optimal matched filters in place of the CTH filters used in previous studies. 

The value of $A_\rmn{ISW}$ that we obtain is larger than the $\Lambda$CDM expectation but consistent with it at $1.2\sigma$, similar to other results using luminous red galaxies in cross-correlation \citep[e.g.][]{Giannantonio:2012}. This is in contrast to the high-significance detections of the stacked ISW signal reported by \citet{Granett:2008a} and \citet{Planck:2015ISW}, which exceed the $\Lambda$CDM expectation by a factor of $\sim5$ or more, corresponding to a $\gtrsim3\sigma$ discrepancy \citep[see][]{Nadathur:2012,Flender:2013,Cai:2014,Hotchkiss:2015a,Aiola:2015}. Such a large discrepancy has been hard to explain in any alternative theoretical models. Our result is therefore an important step towards the resolution of this apparent anomaly.

Our result is also relevant to the proposed explanation of the CMB `Cold Spot' as being due to the ISW effect of a giant void \citep{Szapudi:2015}. This would require a very large enhancement of $A_\rmn{ISW}$ \citep{Nadathur:2014b,MarcosCaballero:2016}, which is not supported by our data.

Finally, it is also noteworthy that the ISW detection method presented here has a sensitivity similar to that of the traditional cross-correlation of projected galaxy density maps with the CMB. For comparison, \citet{Planck:2015ISW} reported an expected $S/N$ of 1.79 for the cross-correlation of the combined SDSS CMASS and LOWZ surveys with \emph{Planck}, albeit based on a data release with smaller sky coverage and photometric redshifts. Our method allows precise measurements of the ISW effect and thus the dynamic effects of dark energy specifically in the extreme density environments of voids and superclusters. It will therefore be useful in further tests of $\Lambda$CDM with future LSS data.

\acknowledgments

We thank Shaun Hotchkiss for useful discussions. SN acknowledges an Individual Fellowship of the Marie Sk\l odowska-Curie Actions under the H2020 Framework of the European Commission, project {\small COSMOVOID}. RC is supported by the UK Science and Technologies Facilities Council grant ST/N000668/1.

This work has made use of public data from the SDSS-III collaboration. Funding for SDSS-III has been provided by the Alfred P. Sloan Foundation, the Participating Institutions, the National Science Foundation, and the U.S. Department of Energy Office of Science. The SDSS-III website is \url{http://www.sdss3.org/}. The Big MultiDark simulations were performed on the SuperMUC supercomputer at the LeibnizRechenzentrum in Munich, using resources awarded to PRACE project number 2012060963. We acknowledge use of the {\small EREBOS}, {\small THEIA} and {\small GERAS} clusters at the Leibniz Institut f\"ur Astrophysik (AIP).

\bibliographystyle{aasjournal}
\bibliography{refs.bib}

\end{document}